\documentclass[doublecol]{epl2} 
\usepackage{graphicx}
\usepackage{amsmath}
\usepackage{amssymb}
\usepackage{url}
\usepackage{color}

\usepackage{ifpdf}
\ifpdf
    \usepackage{epstopdf}

\fi
\usepackage{listings}

\definecolor{gray}{rgb}{0.75,0.75,0.75}

\graphicspath{{.}{../}{./eps/}}

\title{Comment on ``Anderson transition in disordered graphene''}

\author{J.~Schleede\inst{1} \and G.~Schubert\inst{1,2} \and H.~Fehske\inst{1}}
\shortauthor{J.~Schleede \etal}

\institute{                    
  \inst{1} Institut f\"ur Physik, Ernst-Moritz-Arndt-Universit\"at Greifswald - 17489 Greifswald, Germany, EU\\
  \inst{2} Regionales Rechenzentrum Erlangen, Friedrich-Alexander-Universit\"at Erlangen-N\"urnberg - 91058 Erlangen, Germany, EU
}

\date{\today}

\begin{document}

\maketitle

In a recent letter~\cite{Amini2009Anderson}, Amini \textit{et al.} claim 
having found a mobility edge in graphene, a truly two-dimensional (2D) system. 
Their mobility edge ought to be induced by on-site uncorrelated (Anderson-type) 
disorder and---unlike to 3D systems---shall separate 
localized states in the band center, from the remaining extended states. 
In order to distinguish localized 
from extended states, the authors analyzed the  `typical' density 
of states (DOS)~\cite{An58},
\begin{equation}
\rho_\textrm{typ}(E) = 
\exp\left(\frac{1}{K_r K_s}\sum_{r,s}^{K_r,K_s}\ln \rho_{s}^{r}(E)\right)\,,
\label{eq:rho_typ}
\end{equation}
by means of the so-called `Regularized Kernel Polynomial Method' 
(RKPM)~\cite{Sota2007Fast}.

In this comment, we argue that the main conclusion of the above letter
is wrong. Employing the Chebyshev-based Kernel Polynomial Method 
(KPM)~\cite{Voter1996Linearscaling}, we demonstrate that the Anderson model on the graphene lattice, with uniformly distributed on-site energies in the interval $[-W/2,W/2]$, shows localization throughout the whole spectrum, even for small disorder strengths. That is, no mobility edge exists at all.

In particular, we want to make the following points:
(i) The RKPM is not well suited for the evaluation of the typical 
DOS, addressing the subtle 2D Anderson localization 
phenomenon.
(ii) A careful finite-size scaling is necessary to distinguish localized from extended states, either by means of the typical DOS or by considering the whole distribution of the local density of states (LDOS), $\rho_i = \rho_s^r$.

Let us substantiate these assertions in more detail:

(i) The RKPM is a polynomial expansion based on Gaussian-broadened Legendre polynomials $\langle P_n(x) \rangle_\sigma$. 
This broadening is supposed to damp the Gibbs oscillations known from finite order polynomial expansions. 
However, the kernel of this method is not strictly positive (as opposed to the Chebyshev-based KPM with Jackson kernel \cite{Voter1996Linearscaling}), \textit{i.e.}, the approximation of a positive function may become negative in some regions.
\begin{figure}
\includegraphics[width=1.0\columnwidth]{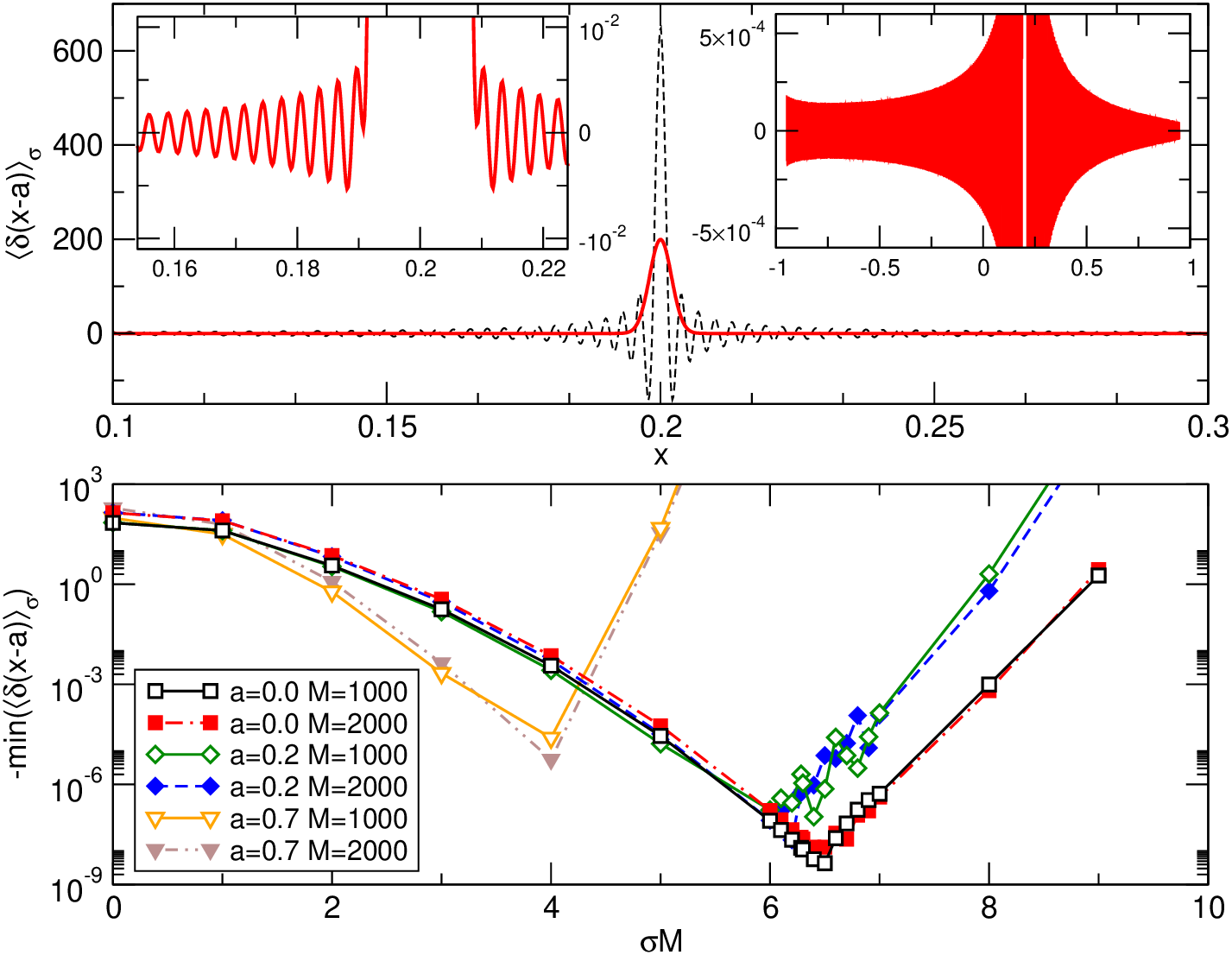}
\caption{(Color on-line) Upper panel: Comparison of pure ($\sigma=0$ -- dashed line) and broadened ($\sigma=4/M$ -- solid red line) expansion of 
a $\delta$-function by Legendre polynomials of order $M=2000$. The two insets show a magnification of the broadened expansion over different ranges. Lower panel: Minimum value of $\langle \delta(x-a)\rangle_\sigma$ for various parameters of $a$, $M$ and $\sigma$.}
\label{fig:wrong_kernel}
\end{figure}
This is illustrated in fig.\ref{fig:wrong_kernel}, which  
gives the Legendre polynomial expansion of a delta-function $\delta(x-a)$ 
(upper main panel). The insets display magnifications of the 
broadened expansion, proving the existence of negative values.
To corroborate this fact we show by the lower panel of fig.~\ref{fig:wrong_kernel} that the expansion of $\delta(x-a)$ contains negative values for various peak positions $a$, expansion orders $M$ and 
Gaussian broadening $\sigma$. 
Although there is an optimal value $\sigma M$, negative values persist for all $\sigma M$.
In addition, the underlying recurrence for the expansion coefficients becomes numerically instable for $\sigma \gtrsim 8/M$ and the expansion characteristics 
strongly depends on the function expanded.

\begin{figure}
\includegraphics[width=1.0\columnwidth]{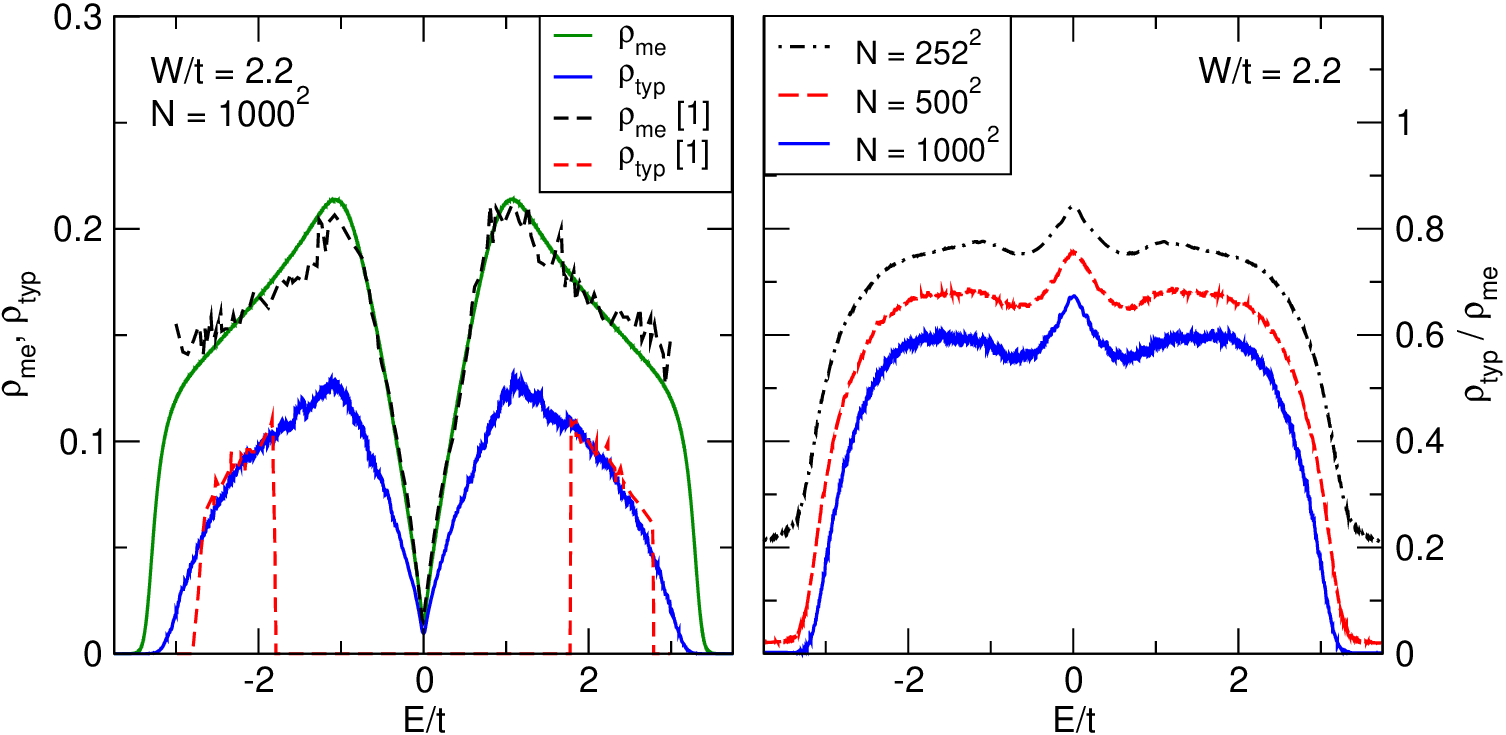}
\caption{(Color on-line) Left panel: Comparison of our KPM results for $\rho_\textrm{typ}$ ($\rho_\textrm{me}$) with RKPM data taken from \cite{Amini2009Anderson}, using identical system size $N=1000^2$, disorder strength $W/t=2.2$ and kernel width $\sigma=8\times10^4$.
Right panel: Finite-size scaling of normalized typical DOS $\rho_\textrm{typ}/\rho_{\textrm{me}}$ at constant resolution ($N_k=500$ states within the kernel). This resolution corresponds to the one of the RKPM data at the maximum of the DOS. 
The total number of realizations are $K_r \times K_s = 61440, 25600, 11712$ for $N=252^2, 500^2, 1000^2$, respectively.
}
\label{fig:me_typ}
\end{figure}

The strict positivity of the kernel is of fundamental 
importance for studying localization by means of the LDOS distribution approach.
Artificial negative values of $\rho_s(E)$ will suppress the typical DOS \eqref{eq:rho_typ} at energies where it would be normally finite.
This issue causes the region of vanishing $\rho_\textrm{typ}$ in the vicinity of the band center for $W/t\geq 1.4$ in fig.~3 of~\cite{Amini2009Anderson}.
To pinpoint this drawback, in the left panel of fig.~\ref{fig:me_typ}, 
we contrast the data of Amini {\it et al.}~\cite{Amini2009Anderson} with our results obtained by KPM with the positive Jackson kernel using matching parameters.
Our data do not show any vanishing $\rho_\textrm{typ}$.
Obviously ,Amini \textit{et al.} wrongly attribute physics to a numerical deficiency.
At this point, we like to emphasize that even for a correct positive kernel, $\rho_\textrm{typ}$ may erroneously vanish if the kernel width becomes comparable to the level spacing.
Since in graphene the mean level spacing strongly varies throughout the spectrum, an adaption of the kernel width to the mean DOS is 
necessary~\cite{Schubert2008Diffusion}.
This guarantees that for any fixed energy at least some energy levels are within the width of the kernel for each realization of disorder.

\begin{figure}
\includegraphics[width=1.0\columnwidth]{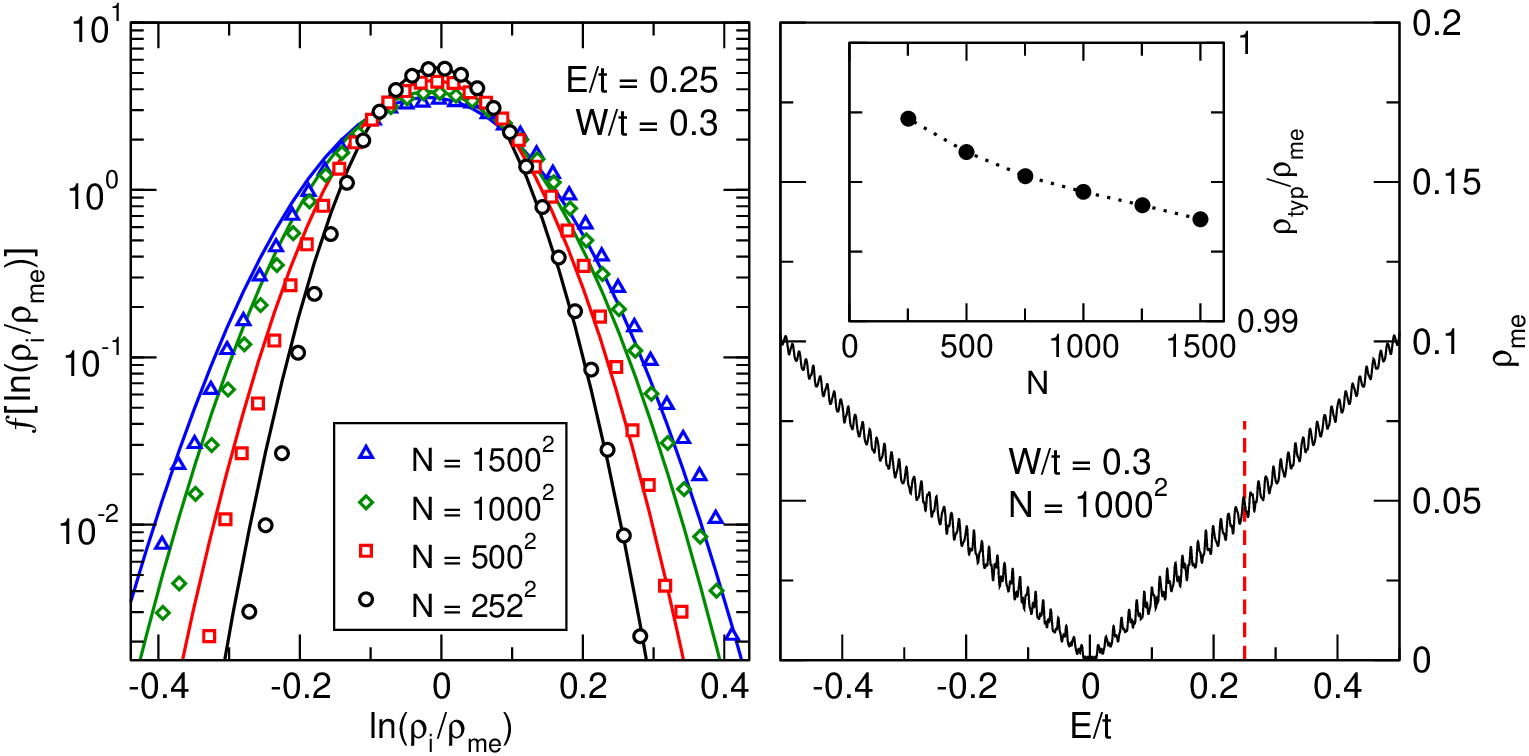}
\caption{(Color on-line) Finite-size scaling of the LDOS distribution with corresponding log-normal fit (left-hand panel) and normalized typical DOS $\rho_\textrm{typ}/\rho_\textrm{me}$ (inset right-hand panel) at energy $E/t = 0.25$ for weak disorder strength $W/t = 0.3$. 
The total number of realizations are $4\times 10^4$ for $N=1500^2$ and $10^5$ otherwise.
}
\label{fig:fss}
\end{figure}

(ii) The use of only a single finite system size to distinguish localized from extended states by means of $\rho_\textrm{typ}$ is insufficient.
In particular, a finite value of $\rho_\textrm{typ}$ does not guarantee extended states.
Instead, the scaling behavior of the typical DOS should be considered (see 
fig.~\ref{fig:me_typ}, right-hand panel).
If $\rho_\textrm{typ}(E)/\rho_\textrm{me}(E)$ remains constant for different system sizes, the corresponding states are extended, whereas $\rho_\textrm{typ}\xrightarrow{N\to\infty}0$ indicates Anderson localization.
According to our results all states in graphene are localized for any 
disorder strength, 
in agreement with the one-parameter scaling theory of localization for 2D lattices \cite{Abrahams1979Scaling}. 
Note that the sensitivity of $\rho_\textrm{typ}$ is limited: 
for localization lengths much larger than the system size, a finite-size scaling of the whole LDOS distribution permits a more reliable detection of the localization properties~\cite{Schubert2008Diffusion}, offering two main advantages. 
First, reducing the distribution characteristics to $\rho_\textrm{typ}$ makes only sense if the shape is close to log-normal.
In more complex systems, this approximation might be poor. 
Second, prior to a pronounced shift of the distribution maximum, \textit{i.e.} a variation of $\rho_\textrm{typ}$, localization is indicated by a broadening of the LDOS distribution with increasing system size (see fig.~\ref{fig:fss}). 
As a consequence, our data for the LDOS distribution clearly indicate 
localization, also for weak disorder.

To conclude, we have shown that under uncorrelated on-site disorder, 
graphene behaves like ordinary 2D systems, which show Anderson localization 
of all single-electron states. Hence, in contradiction with the findings 
by Amini \textit{et al.}~\cite{Amini2009Anderson}, 
Dirac fermions will not remain delocalized below a critical disorder 
strength. The correct analysis of the typical DOS, supplemented by a careful
finite-size scaling, is unsuggestive of mobility 
edges around the Dirac point of graphene.  

\acknowledgments
This work was supported by SPP 1459 of the Deutsche Forschungsgemeinschaft.
%



\begin{thebibliography}{0}
\bibitem{Amini2009Anderson}
\Name{Amini M., Jafari S.~A. \and Shahbazi F.} 
\REVIEW{EPL}{87}{2009}{37002}.

\bibitem{An58}
\Name{Anderson P.~W.}
\REVIEW{Phys. Rev.}{109}{1958}{1498};
\Name{Dobrosavljevi\'{c} V.  \and Kotliar G.}
\REVIEW{Phys. Rev. Lett.}{78}{1997}{3943}.

\bibitem{Sota2007Fast}
\Name{Sota S. \and Itoh M.} \REVIEW{J. Phys. Soc. Jpn.}{76}{2007}{054004}.


\bibitem{Voter1996Linearscaling}
\Name{Voter A.~F., Kress J.~D. \and Silver R.~N.} 
\REVIEW{Phys. Rev. B}{53}{1996}{12733};
\Name{Weisse A., Wellein G., Alvermann A. \and Fehske H.} 
\REVIEW{Rev. Mod. Phys.}{78}{2006}{}; 
\Name{Weisse A. \and Fehske H.} 
\REVIEW{Lect. Notes Physics}{739}{2008}{545}.



\bibitem{Abrahams1979Scaling}
\Name{Abrahams E., Anderson P.~W., Licciardello D.~C. \and Ramakrishnan T.~V.}
\REVIEW{Phys. Rev. Lett.}{42}{1979}{673}.


\bibitem{Schubert2008Diffusion}
\Name{Schubert G. \and Fehske H.}
\REVIEW{Phys. Rev. B}{78}{2008}{155115};
\Name{Schubert G. \and Fehske H.}
\REVIEW{Lect. Notes Physics}{762}{2009}{135};
\Name{Schubert G., Schleede J.,  Byczuk K., Fehske H. \and Vollhardt D.}
\REVIEW{Phys. Rev. B}{81}{2010}{155106}.

\end{thebibliography}
\end{document}